# Giant memory function based on the magnetic field history of resistive switching under a constant bias voltage

Masaya Kaneda*, Shun Tsuruoka, Hikari Shinya, Tetsuya Fukushima, Tatsuro Endo, Yuriko Tadano, Takahito Takeda, Akira Masago, Masaaki Tanaka, Hiroshi Katayama-Yoshida, and Shinobu Ohya**

M. Kaneda, S. Tsuruoka, H. Shinya, T. Endo, Y. Tadano, M. Tanaka, S. Ohya
Department of Electrical Engineering and Information Systems, The University of Tokyo, 7-3-1 Hongo, Bunkyo-ku, Tokyo 113-8656, Japan

H. Shinya, M. Tanaka, H. Katayama-Yoshida, S. Ohya
Center for Spintronics Research Network, Graduate School of Engineering, The University of Tokyo, 7-3-1 Hongo, Bunkyo-ku, Tokyo 113-8656, Japan

H. Shinya, T. Fukushima, A. Masago, H. Katayama-Yoshida
Center for Spintronics Research Network, Graduate School of Engineering Science, Osaka University,1-3 Machikaneyama, Toyonaka, Osaka 560-8531, Japan

H. Shinya
Institute for Chemical Research, Kyoto University, Gokasho, Uji, Kyoto 611-0011, Japan

H. Shinya
Center for Science and Innovation in Spintronics (CSIS), 2-1-1 Katahira, Aoba-ku, Sendai, Miyagi 980-8577, Tohoku University, Japan

T. Fukushima
Research Center for Computational Design of Advanced Functional Materials, National Institute of Advanced Industrial Science and Technology (AIST), 1-1-1 Umezono, Tsukuba, Ibaraki 305-8560, Japan

T. Takeda
Graduate School of Advanced Science and Engineering, Hiroshima University, 1-3-1 Kagamiyama, Higashi-Hiroshima 739-8526, Japan

A. Masago

Research Institute for Value-Added-Information Generation, Japan Agency for Marin-Earth Science and Technology, 3173-25 Showa-machi, Yokohama, Kanagawa 236-0001, Japan

E-mail: *kaneda-masaya@g.ecc.u-tokyo.ac.jp, **ohya@cryst.t.u-tokyo.ac.jp

**Memristors, which are characterized by their unique input-voltage-history-dependent resistance, have garnered significant attention for the exploration of next-generation in-memory computing, reconfigurable logic circuits, and neural networks. Memristors are controlled by the applied input voltage; however, the latent potential of their magnetic field sensitivity for spintronics applications has rarely been explored. In particular, valuable functionalities are expected to be yielded by combining their history dependence and magnetic field response. Here, for the first time, we reveal a giant memory function based on the magnetic field history of memristive switching, with an extremely large magnetoresistance ratio of up to 32,900% under a constant bias voltage, using a two-terminal Ge-channel device with Fe/MgO electrodes. We attribute this behavior to colossal magnetoresistive switching induced by the $d^0$ ferromagnetism of Mg vacancies in the MgO layers and impact ionization breakdown in the Ge substrate. Our findings may lead to the development of highly sensitive multi-field sensors, high-performance magnetic memory, and advanced neuromorphic devices.**

# 1. Introduction

In recent years, the development of the Internet of Things, big data, and artificial intelligence has led to a growing demand for new memory devices with high capacity, low power consumption, and high-speed operation. Memristors are some of the most promising candidates attracting attention.[1] Memristors have been deeply studied, particularly in the fields of in-memory computing[2] and reconfigurable logic.[3,4] Furthermore, their potential in neural networks (NN) has garnered increasing interest.[5–7] Fully integrated memristor-based NN hardware implementations on complementary metal–oxide–semiconductor chips have already been demonstrated.[8,9] Memristors hold promise in NN not only because they can overcome the von Neumann bottleneck through in-memory computing but also because they can efficiently handle multiply-accumulate operations—fundamental and frequently occurring in NN—by processing them as analog signals.

Memristors are characterized by a resistance change that depends on the input voltage history, resulting in hysteresis in the current-voltage characteristic. Memristors are typically two-terminal devices consisting of an insulating layer sandwiched between metal electrodes. Their operation is based on several physical phenomena that occur in the insulating layer.[10] A representative operating principle is resistive switching (RS), in which conductive filaments are formed and ruptured in the insulating layer due to the redistribution of oxygen vacancies or metal atoms under an applied electric field. Other mechanisms include resistance changes due to tunnel magnetoresistance, spin-orbit torque effects,[11] and polarization reversal in ferroelectrics.[12] These mechanisms enable memristors to exhibit resistance changes in response to an applied voltage. Meanwhile, the dependence of memristors on a magnetic field is generally not well known. Detailed studies on the magnetic field dependence of memristors have been limited to a few material systems,[13–23] such as GaAs,[13–14,15,16] BaTiO$_3$/FeMn/BaTiO$_3$,[17] TiO$_2$,[18] and Fe/MgO/Ge.[23] In these systems, the memory bits of memristors cannot be manipulated by the magnetic field history. The hidden potential that might be realized by combining their history dependence and magnetic field response is largely unexplored.

Here, for the first time, we demonstrate a giant memory function based on the magnetic field history of memristive switching using a two-terminal Ge-channel device with Fe/MgO electrodes. In this device, we obtain a memory function with a large magnetoresistance ratio (MR) of up to 32,900% under a constant bias voltage. The very large resistance change, which can be controlled by a magnetic field as well as an electric field, will open a new path for developing highly efficient magnetic memory, susceptible multi-field sensors, and advanced neuromorphic computing devices, which are essential for the next-generation information society.

2. **Giant memory function based on the magnetic field history of resistive switching**

We grew an all-epitaxial single-crystalline heterostructure composed of Co (5 nm)/Fe (17 nm)/MgO (1 nm)/Ge:B (17 nm, B concentration: 1 × 10$^{18}$ cm$^{-3}$)/Ge (51 nm) on an $n^-$–Ge (001) substrate (low $n$-type doping of ~10$^{17}$ cm$^{-3}$) using molecular beam epitaxy (MBE), as illustrated in **Figure** 1a. During growth, the *in-situ* reflection high-energy electron diffraction (RHEED) patterns were streaky, indicating that the Ge:B, MgO, and Fe layers were epitaxially grown on the substrate (Figure S1, Supporting Information). The heterostructure is composed of single-crystal layers with some distortion in the MgO layer, as shown in the cross-sectional lattice image taken by bright field-scanning transmission electron microscopy (bright field-STEM) (Figure 1b). This is further confirmed by high-angle annular dark-field STEM (Figure S2, Supporting Information). This distortion in the MgO layer likely arises from the lattice mismatch between MgO and the underlying Ge:B layer, creating strain in MgO. Such distortion contributes to the formation of Mg vacancies, which play a crucial role in the RS behavior described later. After growth, we partially etched the epitaxially grown layers to make a two-terminal device with drain and source electrodes using electron-beam lithography and Ar-ion milling (Figure 1c). Here, these electrodes are separated by a 3-μm-wide $n^-$-Ge substrate channel. The introduction of the $n^-$-Ge channel is a key structural difference from the one previously reported (Ref. 23). We grounded the source electrode and applied a bias voltage $V$ to the drain electrode to measure the drain-source current $I$ at 3 K. As described below, we obtained the drain-source resistance $R(H) = V/I$ under an external magnetic field $H$. The schematic band diagram of our device in Figure S3, Supporting Information. Here, carriers injected from a Fe electrode transport through the non-doped Ge layer from the valence band of Ge:B to the conduction band of the $n^-$-Ge substrate due to a high electric field and conduct to another Fe electrode.

Two abrupt current jumps appear in the transport characteristics measured at 3 K with increasing $V$ (Figure 1d): one occurs in the low-voltage region of $V$ = 0–10 V and the other in the high-voltage region of $V$ = 10–30 V. Two hysteresis loops corresponding to these jumps can be observed in the $I$–$V$ characteristics. We distinguish the two resistance states in each hysteresis loop, by referring to them as the high-resistance state (HRS) and the low-resistance state (LRS). We define $V_{HRS \to LRS}$ and $V_{LRS \to HRS}$ as the threshold voltages at which a rapid increase and a rapid decrease in $I$ occur, respectively. With increasing $H$ applied along the in-plane [010] direction, these hysteresis loops shift toward higher voltages, as shown by the shift from the purple to orange curves in Figure 1d. Both $V_{\text{HRS}\to\text{LRS}}$ and $V_{\text{LRS}\to\text{HRS}}$ increase with increasing $H$, as indicated by the red and black arrows in Figure 1d. Similar results are obtained for other magnetic-field orientations, such as [$\bar{1}$00] and [$\bar{1}$10], indicating that this system shows no significant magnetic field-direction dependence (Figure S4, Supporting Information).

The behavior observed in the low-voltage region of $V$ = 0–10 V is similar to the colossal magnetoresistive switching (CMRS) observed in a previous report on an Fe/MgO/Ge system[18]: In this low-voltage region, the width of the hysteresis loop, i.e., the difference between $V_{\text{HRS}\to\text{LRS}}$ and $V_{\text{LRS}\to\text{HRS}}$, is small (**Figure** 2a). The hysteresis remains narrow under the application of $H$; with increasing $H$, both $V_{\text{HRS}\to\text{LRS}}$ and $V_{\text{LRS}\to\text{HRS}}$ shift toward higher voltages. When $V$ is fixed while $H$ is swept, our device shows MR curves similar to those in Ref. 23 (Figure 2b).

Meanwhile, the phenomenon observed in the high-voltage region greatly differs. The difference between $V_{\text{HRS}\to\text{LRS}}$ and $V_{\text{LRS}\to\text{HRS}}$ is larger than that in the low-voltage region. The hysteresis is considerably widened by the application of a magnetic field (Figure 1d). As shown in Figure 2c–f, as $H$ increases, the $I$–$V$ curves shift toward higher voltages for both up and down sweeps of $V$, and $V_{\text{HRS}\to\text{LRS}}$ and $V_{\text{LRS}\to\text{HRS}}$ increase with increasing $H$, as in the low-bias region. However, the modulation of $V_{\text{HRS}\to\text{LRS}}$ is greater than that of $V_{\text{LRS}\to\text{HRS}}$; the changes in $V_{\text{HRS}\to\text{LRS}}$ and $V_{\text{LRS}\to\text{HRS}}$ when increasing $H$ from 0 to 5 kG, defined as $\Delta V_{\text{HRS}\to\text{LRS}}$ and $\Delta V_{\text{LRS}\to\text{HRS}}$, are ~10 V and ~5 V, respectively; thus, the hysteresis is enlarged with increasing $H$. This enhancement of the hysteresis with the application of a magnetic field is unique and has never been reported for RS before.

Combining the characteristic magnetic-field-driven changes of the sharp current jumps observed in the high-voltage region (**Figure** 3a) and the $I$–$V$ hysteresis shown in Figure 2c–f, we can demonstrate a giant memory function based on the magnetic field history of RS in this device under a constant bias voltage. As an example, we consider the case when $V$ is fixed at 16.1 V (see the horizontal line in Figure 3a). As shown in Figure 3a, when a large magnetic field of +5 kG is applied, $V_{HRS \to LRS}$ (= 27.0 V) and $V_{LRS \to HRS}$ (= 18.3 V) are greater than 16.1 V, meaning that our device is in the HRS (point A in Figure 3a). Then, with a decrease in $H$ to 0.1 kG (point B), $V_{HRS \to LRS}$ decreases to 16.1 V (see the inset of Figure 3a). Thus, at point B, the resistance state transitions from the HRS to the LRS. Next, with increasing $H$ in the negative direction, the state changes from the LRS to the HRS when $V_{LRS \to HRS}$ increases to 16.1 V (point C). Therefore, we can expect the major loop curve schematically shown in Figure 3b (see the solid line). With increasing $H$ in the positive direction from point B, the state transitions from the LRS to the HRS at point D. This phenomenon is similar to the minor loop behavior of the spin-valve effect observed in conventional spintronics devices, except that point B exists in the $H > 0$ region; however, we can shift point B into the $H < 0$ region by applying a small constant negative magnetic field.

In our experiment, at $V$ = 16.1 V, the device shows spin-valve-like signals, as shown in Figure 3c. We observe a minor loop in the $R$-$H$ characteristics (see the green line in Figure 3c), which represents the historical behavior of the device under a constant voltage. In the major loop, the magnetic field for switching from the LRS to the HRS is different from that at point C, which may be due to the somewhat unstable switching of our device. Memristor devices with a memory function for the magnetic field history have never been reported. The MR ratio obtained in the minor loop measured from 0 kG to 5 kG (green curve in Figure 3c), defined as [$R(H$ = 0.5 kG) $-$ $R(H$ = 0.02 kG)]/$R(H$ = 0.02 kG), is 23,800 %. Furthermore, an even larger MR ratio of 32,900% is obtained at $V$ = 17.0 V (Figure S5, Supporting Information). This value is much larger than the conventional tunnel magnetoresistance (TMR) ratios of magnetic tunnel junctions (at most 1,000%).[24–26] The magnetic-field dependence of the device disappears when the temperature is increased up to 10 K as indicated in Figure S6, Supporting Information.

**3. Possible origins**

In our device, we observe two distinct jumps in $I$ with increasing $V$ and a strong magnetic field dependence of the threshold voltages, as shown in Figure 1d. The resistance state change of memristors is widely attributed to the formation and rupture of conductive filaments (CFs) within their insulating layers. The two distinct current jumps in our device suggest the presence of two separate memristor mechanisms. Each of the two memristor mechanisms discussed below may correspond to one of the two magnetic field-controlled current jumps observed in our device. The first mechanism is RS arising from the electric-field-induced formation of $d^0$ ferromagnetic CFs in the MgO regions. The second mechanism is impact ionization breakdown in the $n^-$–Ge substrate. We note that the space-charge effect[27] and the influence of hydrogen on RS[28] are excluded from our device, as described in ref. 23.

Regarding the first possible mechanism, CFs form in each of the MgO layers (see Figure 1a) when $V$ is applied, resulting in the hysteresis loop in the $I$–$V$ characteristics. RS is often observed in metal oxides, in which the resistance changes due to the formation of filaments caused by the electric-field-induced alignment of vacancies or metal atoms and filament disruption. MgO is a well-known RS material in which both volatile[29,30] and nonvolatile[31,32] RS characteristics have been observed. Additionally, MgO with defects such as Mg vacancies has been experimentally confirmed to exhibit $d^0$ ferromagnetism—ferromagnetism without $d$ orbitals—due to the localized nature of the O 2p orbitals.[33–35] Mg vacancies have local magnetic moments due to onsite Coulomb interactions, and these magnetic moments are aligned by Zener's double exchange interaction, leading to d$^0$ ferromagnetism. [36,37]

Ferromagnetic CFs by Mg vacancies in MgO can explain the observed magnetic field dependence of the threshold voltages. First-principles calculations predict that there is an attractive force between Mg vacancies in MgO, bringing them close together during MgO growth and leading to the partial formation of self-organized *Konbu phase* nanorods.[38] Each Mg vacancy in MgO provides two localized holes in O 2*p* orbitals. Due to the strong onsite Coulomb repulsion, these holes adopt a spin triplet state (**Figure 4a**). When an electric field is applied, Mg vacancies are connected, and the *Konbu phase* is formed, in which hole conduction is induced (Figure 4b). Furthermore, via Zener's double exchange interaction and the magnetic proximity effect from the Fe electrodes, the Mg vacancy spins in MgO tend to align, inducing d$^0$ ferromagnetism (yellow region in Figure 4b). When an external magnetic field is applied, these spins are more strongly aligned; however, at the same time, due to Pauli's exclusion principle, wave function overlap between Mg vacancies is reduced, causing the system to transition into the HRS (Figure 4c), thus increasing the threshold voltage under the magnetic field (see Figure 1d). For the two MgO regions in the current path shown in Figure 1a, the CF formation occurs at nearly the same voltage.

The second possible mechanism is impact ionization breakdown in the $n^-$-Ge substrate. In the $n^-$-Ge substrate, carriers are strongly accelerated by an applied electric field. These accelerated electrons induce impact ionization, by which carriers are generated, in turn causing subsequent impact ionization. This phenomenon further increases the carrier concentration, resulting in a positive feedback loop and breakdown.

Under a strong electric field, carrier flow causes cascading impact ionization, thus forming CFs and leading to memristive behavior (**Figure** 5a). This low-temperature impact ionization breakdown has been observed in semiconductors, such as Si,[39–41] Ge,[42–44] GaAs,[13,14] InSb,[45] and InAs.[46] When a magnetic field is applied, the Lorentz force bends electron trajectories, reducing the effective mean free path. Impact ionization occurs when electrons have enough energy to excite impurity electrons during scattering. In the presence of the magnetic field, the electron energy gained from the electric field between scattering events decreases, which suppresses impact ionization breakdown (Figure 5b). Thus, as the magnetic field increases, a stronger electric field is needed to supply electrons with sufficient energy for impact ionization breakdown.[41,47]

The most plausible scenario is that the two current jumps are independent phenomena; the jump in the low-voltage region originates from RS based on Mg filaments in the MgO layer, while the jump in the high-voltage region is due to impact ionization breakdown in the $n^-$–Ge substrate. As seen in Figure 1d, the observed two current jumps have significantly different characteristics. Specifically, they have different widths of hysteresis and markedly different responses to a magnetic field. The voltage range for the switching observed in the low voltage region is very similar to that observed in a 20-nm Ge channel device with Fe/MgO electrodes in ref. 23, and, thus, this is likely attributed to CF formation in the MgO layer. On the other hand, the significant magnetic field response seen in the high-voltage region, along with the memory function for the magnetic field history shown in Figure 3c, is observed for the first time by using the device with a 3-μm Ge channel and Fe/MgO electrodes shown in Figure 1. The magnetic field response of impact ionization breakdown is known to increase with increasing channel length.[48] Hence, the newly observed memory function may be attributed to impact ionization.

In a reference measurement of the *I–V* characteristics for a *vertical* device made from the same Co/Fe/MgO/Ge:B/Ge/$n^-$–Ge sample, where the voltage is applied between the top Co electrode and the $n^-$–Ge substrate and there is only a single MgO layer in the current path, we observe two current jumps (Figure S7, Supporting Information). This result supports our conclusion that the MgO and Ge layers contribute to the RS. Furthermore, a single hysteresis loop is observed in a lateral device where the Co/Fe/MgO layers were removed from the device shown in Figure 1c (Figure S8, Supporting Information), confirming that the Ge channel can indeed exhibit switching behavior.

To make the observed phenomenon nonvolatile so that the resistance state is maintained even when the voltage is turned off, the formation of more Mg vacancies in the MgO layer may be useful. In this case, the CF formation energy is lowered, which stabilizes the LRS. Thus, we can reduce the required switching voltage, which decreases the power consumption for memory operation. At the same time, the leakage current will be increased, so careful optimization of the MgO layer thickness is necessary. For the Ge regions, the breakdown voltage is influenced by the doping concentration; lower doping concentrations, such as $1 \times 10^{15}$ cm$^{-3}$, result in a higher carrier mobility and a lower breakdown voltage.[43] These strategies focusing on material design and device structure optimization may achieve nonvolatility and low power consumption of the observed memory function in memristive switching.

## 4. Summary

We demonstrate a giant memory function of the magnetic field history of RS controlled by a magnetic field in a two-terminal Fe/MgO/Ge device. This device exhibits an extremely large MR ratio of up to 32,900%, which is two orders of magnitude larger than that of conventional spintronics memory devices, under a constant bias voltage. Our findings reveal the hidden potential of combining the history dependence and magnetic field response of memristors for useful spintronics applications. We propose two possible memristor mechanisms to explain the observed conduction behavior: RS via the $d^0$ ferromagnetic CFs formed in the MgO layers and impact ionization breakdown in the conduction band of the Ge substrate. Our study opens up new avenues for developing highly efficient magnetic memory, susceptible multi-field sensors, and advanced neuromorphic computing devices. The memory function for the magnetic field history of memristive switching under a constant bias voltage, with its high sensitivity to magnetic fields and large MR, has potential for valuable application in information storage, processing, and sensing.

5. **Experimental Section**

We grew an epitaxial single-crystal ferromagnet/insulator/semiconductor structure composed of Co (5 nm)/Fe (17 nm)/MgO (1 nm)/Ge:B (17 nm, B concentration: $1 \times 10^{18}$ cm$^{-3}$)/Ge (51 nm) on an $n^-$-Ge (001) substrate using MBE with a base pressure of $5 \times 10^{-9}$ Pa (Figure 1a). Before growth, the Ge substrate was chemically cleaned with ultrapure water, ammonia water, and acetone, and it was cyclically etched with ultrapure water and buffered hydrogen fluoride for 40 minutes. Then, the substrate was installed in the ultrahigh-vacuum MBE growth chamber through an oil-free load-lock system. The Ge substrate was initially degassed at the substrate temperature $T_S$ = 300 °C for 30 minutes and then thermally cleaned at $T_S$ = 700 °C for 30 minutes. After the thermal treatment, we grew the Ge substrate on the $n^-$-Ge substrate at $T_S$ = 300 °C. The MgO layer was deposited by electron-beam evaporation at $T_S$ = 80 °C with a deposition rate of 0.1 nm/s, followed by the growth of the Fe and Co layers at $T_S$ = 80 °C in the same MBE chamber. After the Fe layer was grown, we annealed the sample at $T_S$ = 200 °C for 10 minutes to improve the flatness of each interface. During growth, the in situ RHEED patterns were streaky, indicating that the MgO and Fe layers were epitaxially grown on Ge:B. The two-terminal device structure with source and drain electrodes was patterned using electron-beam lithography and Ar-ion milling.

**Supporting Information**

Supporting Information is available from the Wiley Online Library of from the author.

**Acknowledgments**

This work is supported by Grants-in-Aid for Scientific Research (No. 24K21214, 22H04948, 20H05650, 23H03802, 23H03805), JST CREST (JPMJCR1777), JST ERATO (JPMJER2202), and Spintronics Research Network of Japan (Spin-RNJ).

**Conflict of Interest**

The authors have no conflicts to disclose.

**Data Availability Statement**

The data that support the findings of this study are available from the corresponding author upon reasonable request.

## Keywords

Keywords: memristors, resistive switching, spintronics

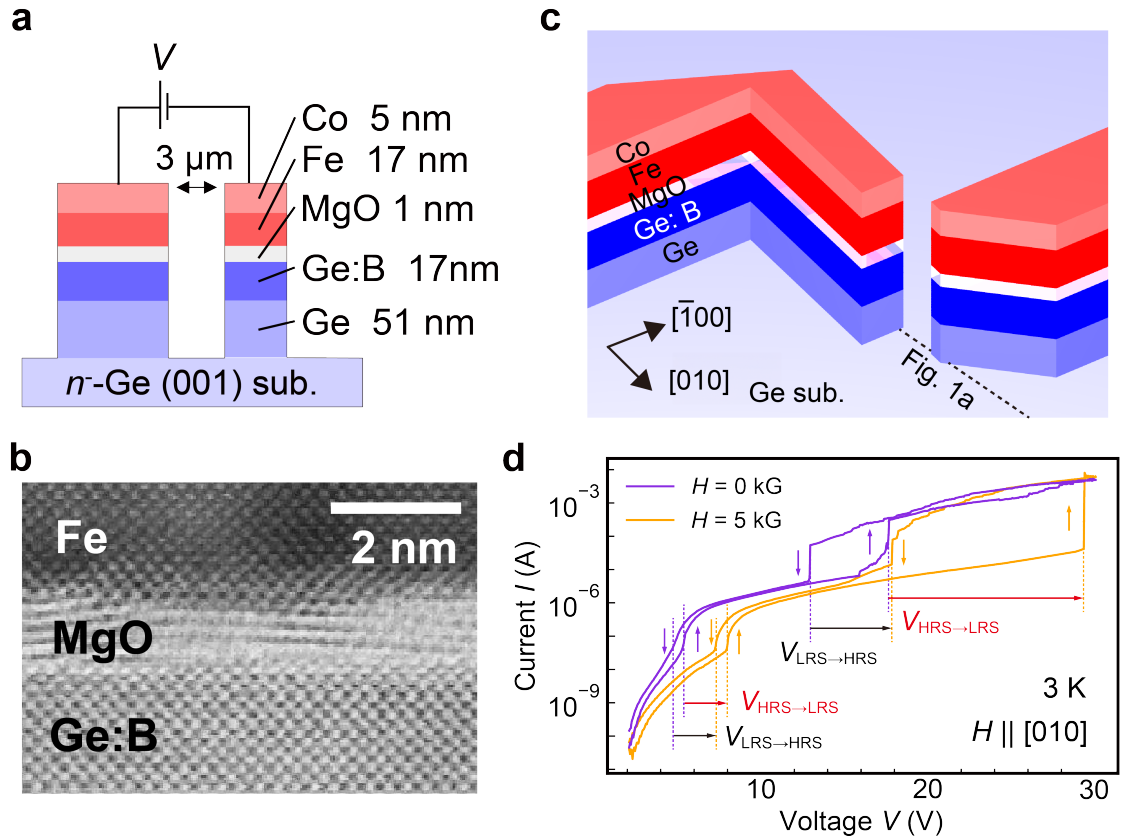

**Figure 1.** Lateral memristive device made from an epitaxial single-crystalline heterostructure composed of Co (5 nm)/Fe (17 nm)/MgO (1 nm)/Ge:B (17 nm; B concentration of $1\times10^{18}$ cm$^{-3}$)/Ge (51 nm) on an $n^-$-Ge (001) substrate. a) Schematic side view of our device. b) Cross-sectional lattice image of the grown sample taken via scanning transmission electron microscopy with the electron beam azimuth along the [100] axis of Ge. c) Schematic top view near the channel region of the sample. The dotted line corresponds to the position of the cross-section shown in a). d) $I$–$V$ characteristics of the device at $H = 0$ (purple) and 5 kG (orange). The arrows represent the voltage $V$ sweep directions. The red and black arrows correspond to the magnetic field-induced voltage shifts of $V_{\text{HRS}\rightarrow\text{LRS}}$ and $V_{\text{LRS}\rightarrow\text{HRS}}$, respectively.

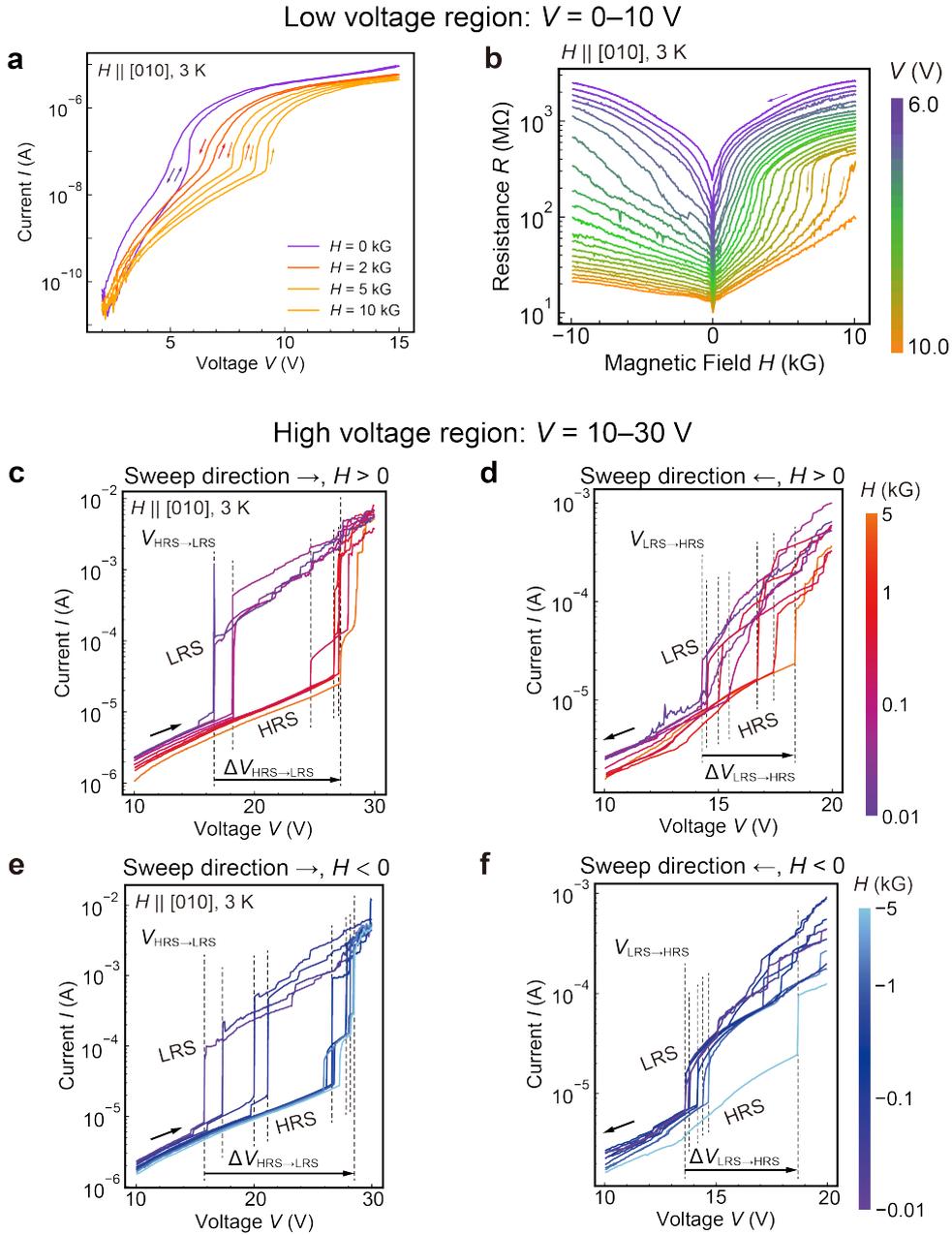

**Figure 2.** Operation of the memristor device. a) *I–V* characteristics for various *H* values at 3 K around the current jumps in the *V* range from 0 to 15 V. The arrows represent the *V* sweep directions. b) *R–H* characteristics when sweeping *H* from 10 to –10 kG in the *V* range from 6 to 10 V (low voltage region). c–f) *I–V* characteristics at various positive (c, d) and negative (e, f) *H* values when *V* is swept up (c, e) and down (d, f) in the range of 10 – 30 V.

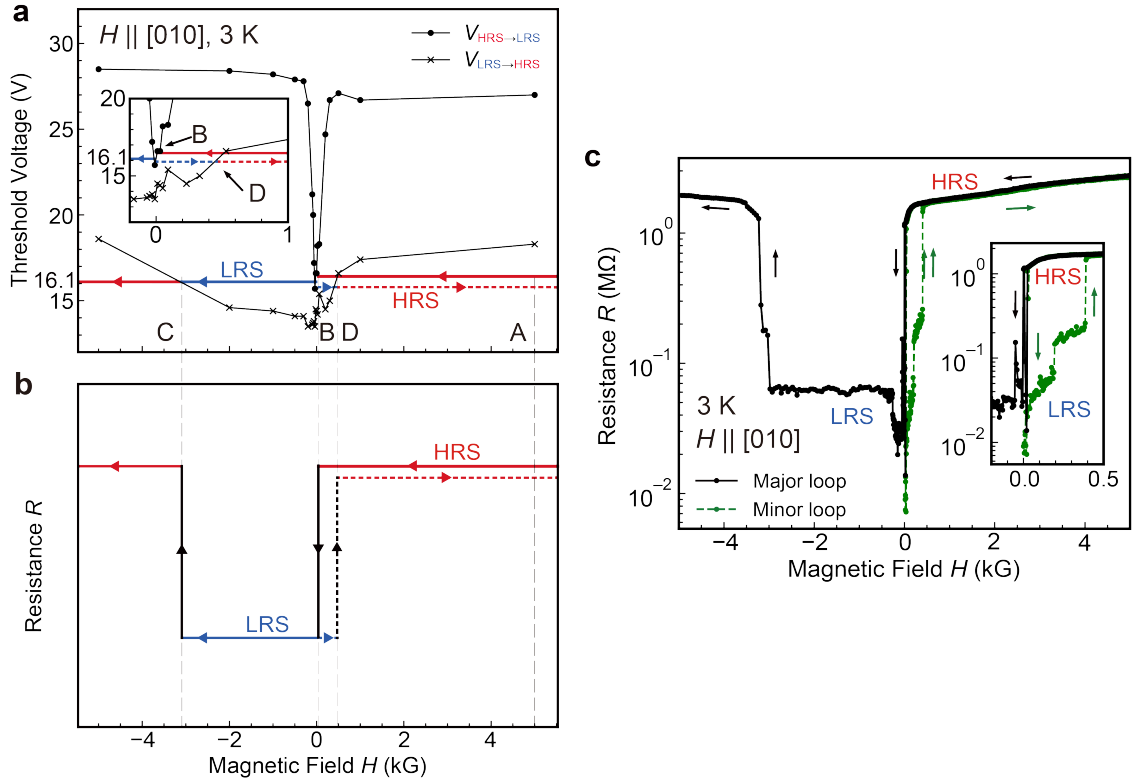

**Figure 3.** Magnetic field dependence of the resistance. a) $V_{\text{HRS}\rightarrow\text{LRS}}$ and $V_{\text{LRS}\rightarrow\text{HRS}}$ obtained in the high-bias voltage region (10 – 30 V) as a function of $H$ (see Fig. 2c–f). b) $R$–$H$ characteristics predicted from (a). c) Major (black) and minor (green) MR loops obtained for the device at 3 K when $V$ = 16.1 V. For those measurements, the magnetic field was applied along the [010] direction.

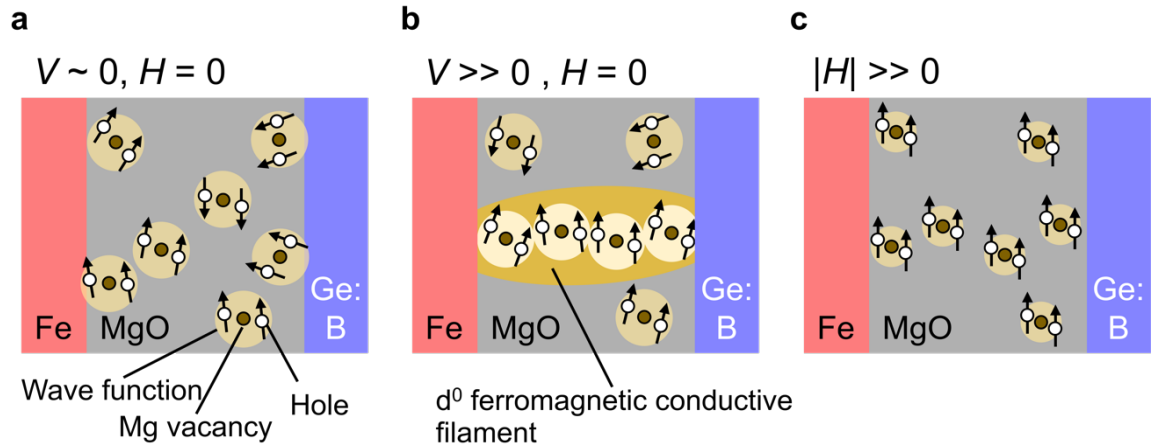

**Figure 4.** Schematic illustration of the RS conduction mechanism in MgO. a-c) Conduction mechanism due to RS in MgO. The brown and white circles represent Mg vacancies and holes, respectively. The arrows express spins. The pale and dark yellow regions represent the hole wave functions and a CF, respectively. a) shows the case under no magnetic field with a low bias voltage, in which Mg vacancies are not aligned, and the device is in the HRS. The Mg vacancies are paramagnetic. b) shows the case when a sufficient bias voltage is applied. The Mg vacancies are aligned, and CFs are formed. The magnetic moments in the CFs are roughly aligned via Zener's double exchange interaction, and the CFs become ferromagnetic. c) shows the case when a strong external magnetic field is applied. The magnetic moments of the CFs are strongly aligned, and the hole wave functions contract due to Pauli's exclusion principle, causing the CFs to collapse.

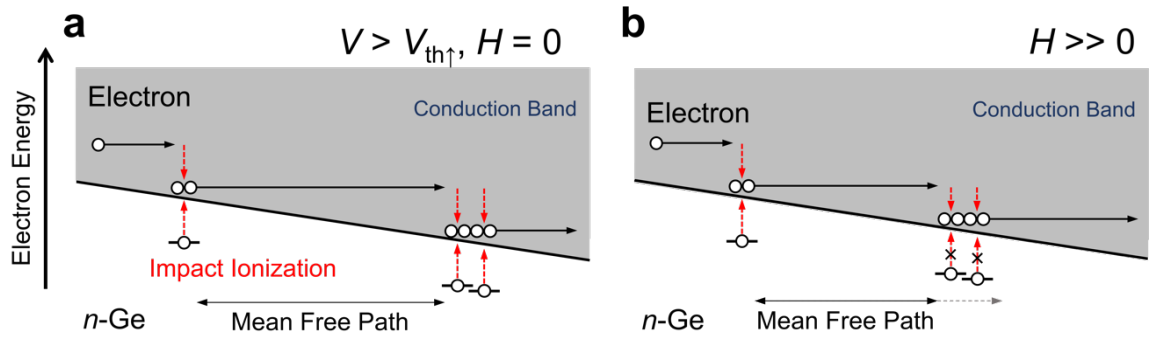

**Figure 5.** a),b) Schematic illustration of the impact ionization conduction mechanism in the $n^-$-Ge substrate. The small circles are electrons. The solid line represents the conduction band edge. The gray and white regions are the conduction band and the band gap, respectively. a) shows hot electrons colliding with impurities, exciting further electrons. The excited electrons are accelerated by the electric field over their mean free path and then generate more electrons through impact ionization. This carrier generation occurs in a cascading manner, leading to ionization breakdown. b) shows the case under a magnetic field, in which the mean free path decreases due to the application of a magnetic field. The energy gained by electrons from the electric field during their mean free path is reduced, and impact ionization does not occur.

Supporting Information

**Giant memory function for the magnetic field history of resistive switching under a constant bias voltage**

Masaya Kaneda, Shun Tsuruoka, Hikari Shinya, Tetsuya Fukushima, Tatsuro Endo, Yuriko Tadano, Takahito Takeda, Akira Masago, Masaaki Tanaka, Hiroshi Katayama-Yoshida, and Shinobu Ohya.

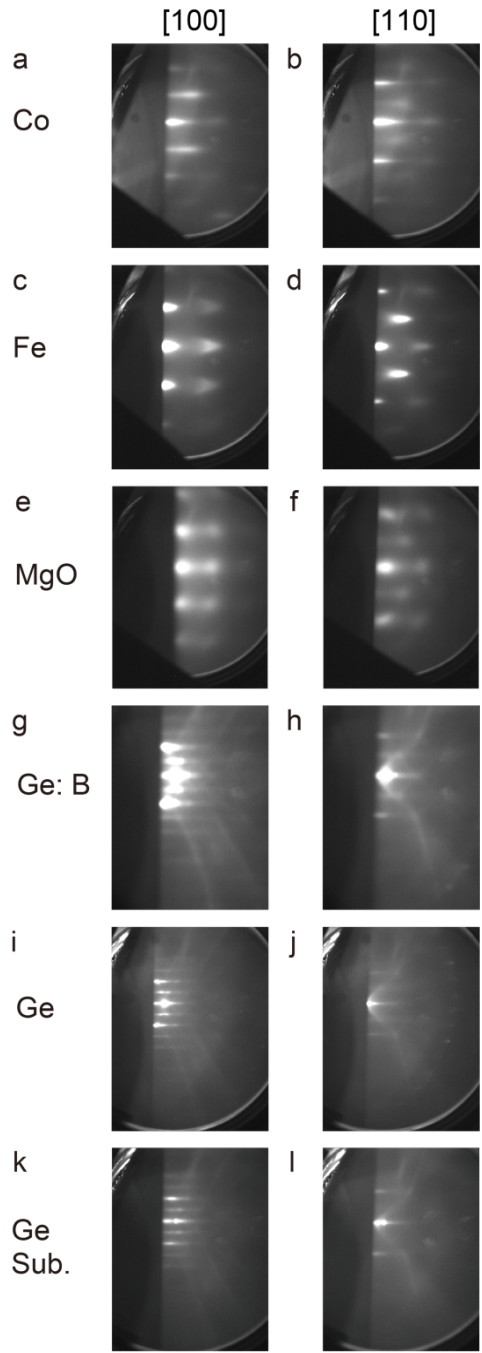

**Figure S1.** Reflection high energy electron diffraction (RHEED) patterns during the growth. RHEED pattern of a),b) Co, c),d) Fe, e),f), MgO, g),h) Ge: B, i),j), Ge and k),l) Ge substrate along the [100] direction (a, c, e, g, i, k) and the [110] direction (b, d, f, h, j, l) of the Ge substrate.

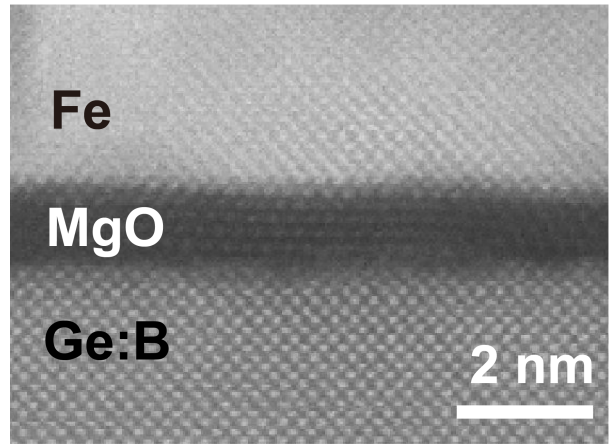

**Figure S2.** High-angle annular dark-field scanning transmission electron microscopy (HAADF-STEM) lattice image of the Fe/MgO/Ge:B layers.

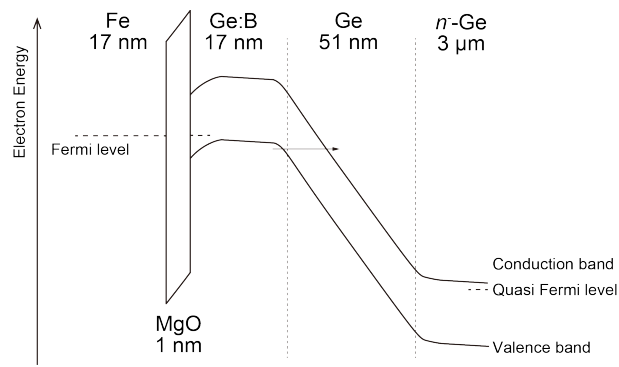

**Figure S3**. Schematic band diagram of the Fe (17 nm)/MgO (1 nm)/Ge:B (17 nm; B concentration of $1\times10^{18}$ cm$^{-3}$)/Ge (51 nm) on an $n^-$-Ge (001) substrate.

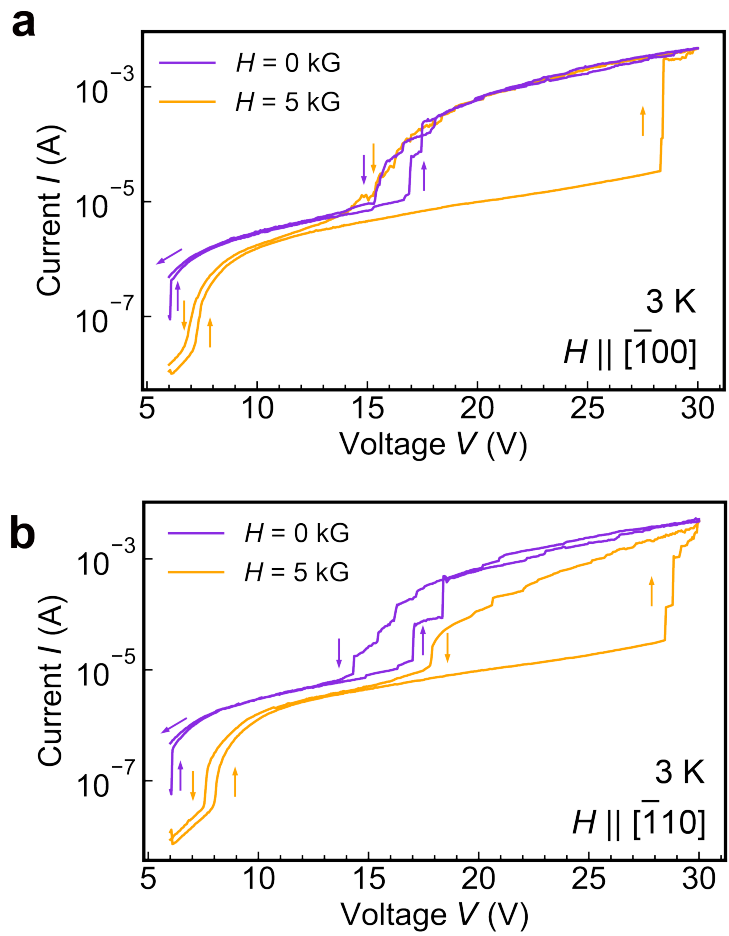

**Figure S4.** *I–V* characteristics of the device at *H* = 0 (purple) and 5 kG (orange) a) when *H* is applied along the [$\bar{1}$00] direction in the film plane and b) when *H* is applied along the [$\bar{1}$10] direction in the film plane.

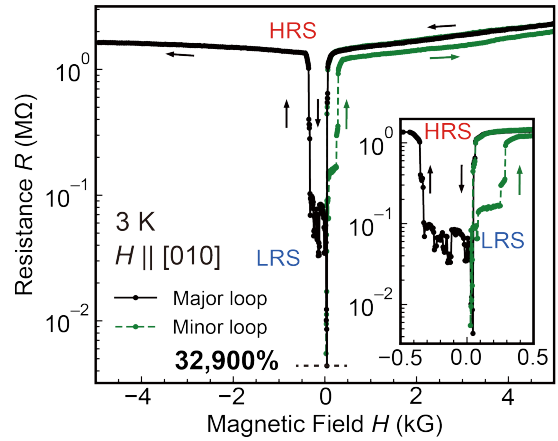

**Figure S5**. Major (black) and minor (green) MR loops obtained for the device at 3 K when $V = 17.0$ V. For those measurements, the magnetic field was applied along the [010] direction. The MR ratio, calculated from the black curve as $[R(H = 0.5$ kG$) – R(H = 0.045$ kG$)]/R(H = 0.045$ kG$)$ is 32,900%.

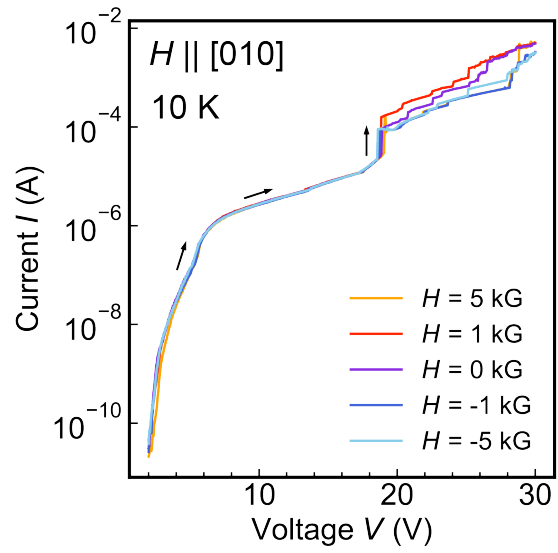

**Figure S6.** *I–V* characteristics of the device at various *H* values at 10 K when $V_{DS}$ increases from zero. *H* applied along the [010] direction in the film plane.

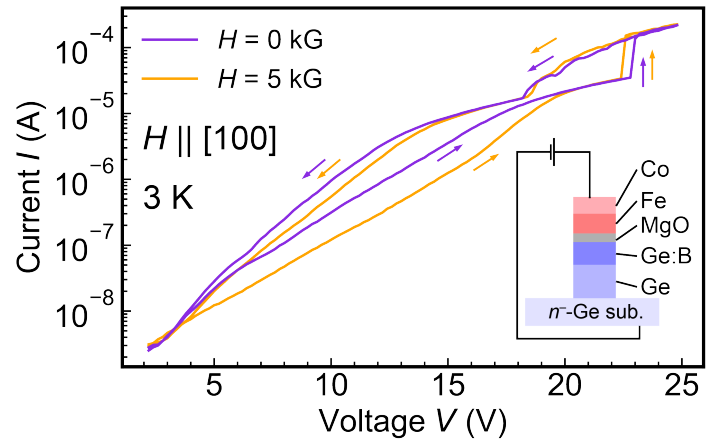

**Figure S7.** *I–V* characteristics of a *vertical* device made from the same Ge:B (17 nm; B concentration of $1\times10^{18}$ cm$^{-3}$)/Ge (51 nm)/$n^-$-Ge (001) heterostructure, in which the current between the top Co electrode and the $n^-$–Ge substrate is measured, at $H = 0$ (purple) and 5 kG (orange).

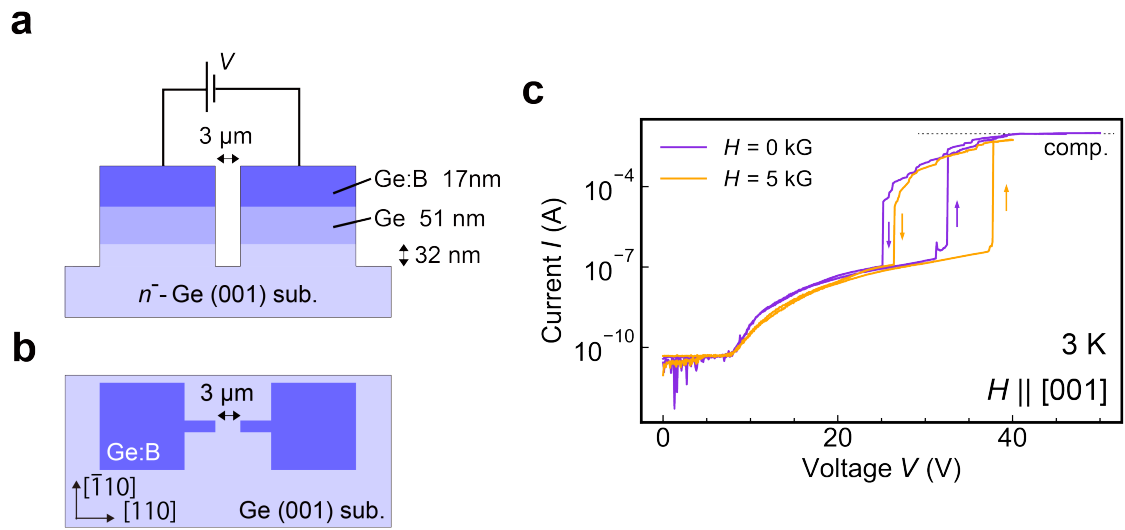

**Figure S8.** Lateral memristive device made from an epitaxial single-crystalline heterostructure composed of Ge:B (17 nm; B concentration of $1\times10^{18}$ cm$^{-3}$)/Ge (51 nm) on an $n^-$-Ge (001) substrate. a),b) Schematic a) side and b) top views of the device. c) *I–V* characteristics of this device at $H = 0$ (purple) and 5 kG (orange).